\pdfoutput=1
\RequirePackage{ifpdf}
\ifpdf 
\documentclass[pdftex]{sigma}
\else
\documentclass{sigma}
\fi

\begin{document}

\newcommand{\arXivNumber}{1503.08669}

\allowdisplaybreaks

\renewcommand{\thefootnote}{$\star$}

\renewcommand{\PaperNumber}{065}

\FirstPageHeading

\ShortArticleName{s2s-OVCA and Flow-Density Relation}

\ArticleName{A CA Hybrid of the Slow-to-Start and the Optimal\\
  Velocity Models and its Flow-Density Relation\footnote{This paper is a~contribution to the Special Issue on Exact Solvability and Symmetry Avatars
in honour of Luc Vinet.
The full collection is available at
\href{http://www.emis.de/journals/SIGMA/ESSA2014.html}{http://www.emis.de/journals/SIGMA/ESSA2014.html}}}

\Author{Hideaki UJINO~$^\dag$ and Tetsu YAJIMA~$^\ddag$}

\AuthorNameForHeading{H.~Ujino and T.~Yajima}

\Address{$^\dag$~National Institute of Technology, Gunma College, Maebashi,
Gunma 371--8530, Japan}
\EmailD{\href{mailto:ujino@nat.gunma-ct.ac.jp}{ujino@nat.gunma-ct.ac.jp}}

\Address{$^\ddag$~Department of Information Systems Science,
Graduate School of Engineering,\\
\hphantom{$^\ddag$}~Utsunomiya University, Utsunomiya 321--8585, Japan}
\EmailD{\href{mailto:yajimat@is.utsunomiya-u.ac.jp}{yajimat@is.utsunomiya-u.ac.jp}}

\ArticleDates{Received March 31, 2015, in f\/inal form July 27, 2015; Published online July 31, 2015}

\Abstract{The s2s-OVCA is a cellular automaton (CA) hybrid
of the optimal velocity (OV) model and the slow-to-start (s2s) model,
which is introduced in the framework of the ultradiscretization method.
Inverse ultradiscretization as well as the time continuous limit,
which lead the s2s-OVCA to an integral-dif\/ferential equation,
are presented.
Several traf\/f\/ic phases such as a free f\/low as well as slow f\/lows
corresponding to multiple metastable states
are observed in the f\/low-density relations of the s2s-OVCA.
Based on the properties of the stationary f\/low of the s2s-OVCA,
the formulas for the f\/low-density relations are derived.}


\Classification{39A10; 39A06}

\renewcommand{\thefootnote}{\arabic{footnote}}
\setcounter{footnote}{0}

\section{Introduction}

Self-driven many-particle systems have provided
a good microscopic point of view on the vehicle
traf\/f\/ic~\cite{Chowdhury2000,Helbing2001}.
The optimal velocity model~\cite{Bando1995}
gives a description of such a system with
a set of ordinary dif\/ferential equations (ODE).
It is a car-following model describing an adaptation to
the optimal velocity that depends on the distance
from the vehicle ahead. Another way of describing such systems is
provided by cellular automata (CA). For example,
the elementary CA of Rule 184 (ECA184)~\cite{Wolfram1986},
the Fukui--Ishibashi (FI) model~\cite{Fukui1996} and the slow-to-start (s2s)
model~\cite{Takayasu1993} are CA describing
vehicle traf\/f\/ic as self-driven many-particle systems.

Studies of the self-driven many-particle systems have been wanting
a framework that commands a bird's eye view of both ODE and CA
models in a unif\/ied manner. Ultradiscretization~\cite{Tokihiro1996},
which gives a link between the Korteweg--de Vries (KdV) equation and
integrable soliton CA~\cite{Takahashi1990},
is expected to provide such a framework, for
it can be applied to non-integrable systems, too.
As a f\/irst step toward such a framework,
an ultradiscretization of the OV model~\cite{Takahashi2009}
was presented.
A specif\/ic choice of the OV function enabled the ultradiscretization of the
OV model without any other specialization.
We should note that another
ultradiscretization of the OV model with essentially the same OV function
as above
was derived from the modif\/ied KdV (mKdV) equation, which is an
ef\/fective theory around the critical point,
with specializing its solutions
to traveling wave solutions~\cite{Kanai2009}. The latter ultradiscretization
of the OV model depends on the ultradiscretization of the mKdV equation,
which is an integrable soliton equation with rich accumulation of
the studies of integrable discretization and ultradiscretization.
The former one, on the other hand, has nothing to do with integrability,
which indicates a possibility to expand the scope of the ultradiscretization
beyond the integrable models. Thus the two ultradiscretizations
make a clear contrast regarding to integrability.
An early search for a CA-type OV model dates back to
1999~\cite{Helbing2001,Helbing1999},
which was done from a phenomenological point of view to highway traf\/f\/ic.

The former ultradiscretization of the OV model~\cite{Takahashi2009}
provided a foundation toward a hybri\-dization of
the OV model and the s2s model,
and it lead to the s2s-OVCA~\cite{Oguma2009} indeed,
without the aid of the integrable
models.
The s2s-OVCA is a CA-type hybrid of the OV model and the s2s model.
As we shall see, the equation of the s2s-OVCA generally involves
three or more times,
or higher order time-dif\/ferences, in other words. As far as the authors know,
dif\/ference equations involving higher order time-dif\/ferences yet want
a thorough study from a point of view of the ultradiscretization.
Besides an interest from the traf\/f\/ic theoretical point of view,
an interest toward a new horizon of the scope of the ultradiscretization
motivates us to introduce and study the s2s-OVCA.
As we shall show in Section~\ref{sec:2},
the s2s-OVCA reduces to an ODE that is an extension of the OV model
in the inverse-ultradiscrete and the time-continuous limits.

It was observed by numerical experiments
that motion of the vehicles described by
the s2s-OVCA went to stationary f\/low in the long run, irrespectively
of the initial conf\/iguration~\cite{Oguma2009,Ujino2012}.
It was also observed by numerical experiments that
the f\/low-density relation for the stationary f\/low of the s2s-OVCA
was piecewise linear and f\/lipped-$\lambda$ shaped with
several metastable slow branches~\cite{Oguma2009}.
Exact expression for the f\/low-density relation was given by
a set of exact solutions giving stationary f\/lows of
the s2s-OVCA~\cite{Ujino2012}.
The f\/lipped-$\lambda$
shaped diagram captures the characteristic of
observed f\/low-density relations~\cite{Chowdhury2000,Helbing2001}.
Some other CA type models that gave a~f\/lipped-$\lambda$ shaped f\/low-density
relation with a metastable branch was also reported~\cite{Barlovic1998}.
We shall explain in Section~\ref{sec:3}
the f\/low-density relation of the s2s-OVCA based on
the properties of the stationary f\/low
which was numerically observed~\cite{Oguma2009}.

\section{s2s-OVCA and its inverse ultradiscretization}
\label{sec:2}

The s2s-OVCA is given by a set of dif\/ference
equations below
\begin{gather}
  x_k^{n+1}=x_k^n+\min\Bigl(\min_{n^\prime=0}^{n_0}\bigl(
  x_{k+1}^{n-n^\prime}-x_{k}^{n-n^\prime}-1\bigr),v_0\Bigr),
  \label{eq:s2s-OVCA2}
\end{gather}
where the integers $n_0\geq 0$, $v_0\geq 0$ and
$x_k^n$, $k=1,2,\dots, K$, are the monitoring period, the top speed and
the position of the car $k$ at the $n$-th discrete time.
Note that the def\/inition of the symbol $\min\limits_{k=0}^{N}$ is
\begin{gather*}
  \min_{k=0}^{N}(a_k):=\min(a_0,a_1,a_2,\dots,a_N).
\end{gather*}
The equation \eqref{eq:s2s-OVCA2} is called
an ultra-discrete equation in the sense that it is a
dif\/ference equation which
is piecewise linear with respect to the dependent variables~$x_k^n$.
The s2s-OVCA includes the ECA184 ($n_0=0$, $v_0=1$)~\cite{Wolfram1986},
the FI model ($n_0=0$)~\cite{Fukui1996} and
the s2s model ($n_0=1$, $v_0=1$)~\cite{Takayasu1993}
as its special cases.

Since the second term in the right hand side of equation~\eqref{eq:s2s-OVCA2}
gives the speed of the car~$k$ at the time~$n$,
the s2s-OVCA describes
many cars running on a single lane highway in one direction, which is driven
by cautious drivers requiring enough headway to go on at least for~$n_0$ time steps before they accelerate their cars.
The equation~\eqref{eq:s2s-OVCA2} also means that the car slows
down immediately when its
headway becomes less than its velocity. Thus the monitoring pe\-riod~$n_0$
describes asymmetry between acceleration and deceleration of the cars.
It is said that the acceleration times are about f\/ive to ten times larger
than the braking times~\cite{Helbing2001}.

Without loss of generality, we can assume that the cars are arrayed
in numerical order, $x_1^0<x_2^0<\cdots<x_K^0$, which is also assumed
throughout below.
Then the number of empty cells between the cars $k$ and $k+1$ for any $k$
is always non-negative, i.e.,
\begin{gather}
  \label{eq:no_clash}
  x_{k+1}^n-x_k^n-1:=\Delta x_k^n-1\geq 0.
\end{gather}
It is obvious that the inequality holds for $n=0$. We assume that
the inequality holds up to some $n$, as the induction hypothesis.
The induction hypothesis as well as the def\/inition of $\min$
assure the inequality
\begin{gather}
  0\leq \min\Bigl(\min_{n^\prime=0}^{n_0}\bigl(
  \Delta x_{k}^{n-n^\prime}-1\bigr),v_0\Bigr)\leq
  \Delta x_{k}^n-1
  \label{eq:2-1}
\end{gather}
for any $k$. Using equation~\eqref{eq:s2s-OVCA2}, we get an expression
of $\Delta x_k^n$ as
\begin{gather}
  \Delta x_k^{n+1}-1  = \Delta x_k^n-1
  +\min\Bigl(\min_{n^\prime=0}^{n_0}\bigl(
  \Delta x_{k+1}^{n-n^\prime}-1\bigr),v_0\Bigr)
  -\min\Bigl(\min_{n^\prime=0}^{n_0}\bigl(
  \Delta x_{k}^{n-n^\prime}-1\bigr),v_0\Bigr) \nonumber\\
\hphantom{\Delta x_k^{n+1}-1}{}
    = \min\Bigl(\min_{n^\prime=0}^{n_0}\bigl(
  \Delta x_{k+1}^{n-n^\prime}-1\bigr),v_0\Bigr)
  +\Bigl[\Delta x_k^n-1
  -\min\Bigl(\min_{n^\prime=0}^{n_0}\bigl(
  \Delta x_{k}^{n-n^\prime}-1\bigr),v_0\Bigr)\Bigr].\label{eq:2-2}
\end{gather}
The inequality \eqref{eq:2-1} and the equation~\eqref{eq:2-2} show that
the inequality~\eqref{eq:no_clash} holds for~$n+1$.
The inequality~\eqref{eq:no_clash} means
that both overtake and clash are prohibited by the s2s-OVCA.

We should note that the s2s-OVCA is obtained from a dif\/ference equation
by a limiting procedure named ultradiscretization~\cite{Tokihiro1996},
which generates
a piecewise-linear equation from a~dif\/ference equation via the limit formula
\begin{gather}
  \label{eq:2-3}
  \lim_{\delta x\rightarrow +0}\delta x \log\biggl(
  \sum_{k=0}^N b_k{\rm e}^{a_k/\delta x}
  \biggr)=\max(a_0,a_1,a_2,\dots,a_N)=:\max_{k=0}^N(a_k),
\end{gather}
where arbitrary numbers $b_k$ must be positive.
The equation~\eqref{eq:2-3} is rewritten as
\begin{gather*}
  \lim_{\delta x\rightarrow +0}\delta x \log\biggl(
  \sum_{k=0}^N b_k{\rm e}^{-a_k/\delta x}
  \biggr)^{-1}=\min_{k=0}^N(a_k),
\end{gather*}
for $\min(a_0,a_1,a_2,\dots,a_N)=-\max(-a_0,-a_1,-a_2,\dots,-a_N)$.

For the sake of convenience in the calculation below, we introduce two
parameters, $x_0$~and~$\delta t$, in the s2s-OVCA
\begin{gather}
  x_k^{n+1}=x_k^n+\min\Bigl(\min_{n^\prime=0}^{n_0}\bigl(
  \Delta x_k^{n-n^\prime}\bigr)-x_0,v_0\delta t\Bigr)
  =:x_k^n+v_{\rm opt}^{\rm u}(\Delta_{\rm ef\/f}x_k^n)\delta t,
  \label{eq:s2s-OVCA}
\end{gather}
where $\Delta_{\rm ef\/f}x_k^n:=\min_{n^\prime=0}^{n_0}
\bigl(\Delta x_k^{n-n^\prime}\bigr)$. The parameters $x_0$ and $\delta t$
are the length of a cell, which corresponds to the space occupied
by a single car or the length of a car itself in the shortest case imaginable,
and the discrete time-step, respectively.
For simulation of highway traf\/f\/ics,
the length of a cell $x_0$ and the discrete time-step
$\delta t$ are
usually chosen as $x_0=7.5$~m and
$v_0=5\times\frac{x_0}{\delta t}$~\cite{Helbing2001}.
But we do not consider specif\/ic values of the parameters in the
s2s-OVCA and regard them as generic.
The two parameters $x_0$ and $\delta t$ were set to be unity in
equation \eqref{eq:s2s-OVCA2}.
Introduction of $x_0$ into the inequality \eqref{eq:2-1} gives
\begin{gather}
  \label{eq:2-5}
  \Delta x_k^n-x_0\geq 0
\end{gather}
for any $n$ and $k$, which is shown by induction with the aid of
an inequality
\begin{gather*}
  0\leq \min\Bigl(\min_{n^\prime=0}^{n_0}\bigl(
  \Delta x_{k}^{n-n^\prime}-x_0\bigr),v_0\delta t\Bigr)\leq
  \Delta x_{k}^n-x_0
\end{gather*}
and an expression of $\Delta x_k^n$ derived from equation~\eqref{eq:s2s-OVCA}
\begin{gather*}
  \Delta x_k^{n+1}-x_0
  = \min\Bigl(\min_{n^\prime=0}^{n_0}\bigl(
  \Delta x_{k+1}^{n-n^\prime}-x_0\bigr),v_0\Bigr)
  +\Bigl[\Delta x_k^n-x_0
  -\min\Bigl(\min_{n^\prime=0}^{n_0}\bigl(
  \Delta x_{k}^{n-n^\prime}-x_0\bigr),v_0\Bigr)\Bigr]
\end{gather*}
that correspond equations \eqref{eq:2-1} and \eqref{eq:2-2}, respectively.

Since we have the inequality~\eqref{eq:2-5} for the headway
$\Delta x_k^n-x_0$, the
ef\/fective headway $\Delta_{\rm ef\/f}x_k^n-x_0$ is also always non-negative,
$\Delta_{\rm ef\/f}x_k^n-x_0\geq 0$, for any~$k$. With the aid of the identity
\begin{gather*}
  \min(A, B)=A-\max(0, A-B)=\max(0, A)-\max(0, A-B)
\end{gather*}
for any $A\geq 0$, the optimal velocity function
$v_{\rm opt}^{\rm u}(x)\delta t:=\min(x-x_0,v_0\delta t)$ in the s2s-OVCA
is expressed as
\begin{gather*}
  v_{\rm opt}^{\rm u}(x)\delta t=\max(0,x-x_0)
  -\max(0,x-x_0-v_0\delta t),
\end{gather*}
for any $x>0$. It is given by the ultradiscrete
limit $\delta x\rightarrow +0$ of a function
\begin{gather*}
  v_{\rm opt}^{\rm d}(x)\delta t=\delta x\log
  \left[\frac{\frac{1+{\rm e}^{(x-x_0)/\delta x}}{1+{\rm e}^{-x_0/\delta x}}}
  {\frac{1+{\rm e}^{(x-x_0-v_0\delta t)/\delta x}}
  {1+{\rm e}^{-(x_0+v_0\delta t)/\delta x}}}
  \right],
\end{gather*}
which is an inverse-ultradiscretization of the optimal
velocity function $v_{\rm opt}^{\rm u}$.
Note that we have introduced arbitrary coef\/f\/icients
so as to make $v_{\rm opt}^{\rm d}(0)=0$.
In a similar way to the above calculation,
an inverse-ultradiscretization of the ef\/fective interval
$\Delta_{\rm ef\/f}^{\rm u}x_k^n$ is also obtained as
\begin{gather*}
  \Delta_{\rm ef\/f}^{\rm d}x_k^n:=\delta x\log
  \left(
    \sum_{n^\prime=0}^{n_0}
    \frac{{\rm e}^{-\Delta x_k^{n-n^\prime}/\delta x}}{n_0+1}
  \right)^{-1}.
\end{gather*}
Therefore an inverse-ultradiscretization of the us2s-OVCA is given by
$x_k^{n+1}=x_k^n+v_{\rm opt}^{\rm d}(\Delta_{\rm ef\/f}^{\rm d}x_k^n)\delta t$,
which is explicitly written as
\begin{gather}
  x_k^{n+1} =x_k^n+\delta x\Biggl\{
  \log\biggl[
      1+\biggl(\sum_{n^\prime=0}^{n_0}
      \dfrac{{\rm e}^{-(\Delta x_k^{n-n^\prime}-x_0)/\delta x}}{n_0+1}
      \biggr)^{-1}\biggr]
  -\log\bigl(1+{\rm e}^{-x_0/\delta x}\bigr) \nonumber\\
\hphantom{x_k^{n+1} =}{} -
  \log\biggl[
      1+\biggl(\sum_{n^\prime=0}^{n_0}
      \dfrac{{\rm e}^{-(\Delta x_k^{n-n^\prime}-x_0-v_0\delta t)/\delta x}}{n_0+1}
      \biggr)^{-1}\biggr]
  +\log\bigl(1+{\rm e}^{-(x_0+v_0\delta t)/\delta x}\bigr)
  \biggr\}.
  \label{eq:ds2s-OV}
\end{gather}
In other words, the s2s-OVCA is given by the ultradiscrete limit
$\delta x\rightarrow +0$ of the above dif\/ference equation~\eqref{eq:ds2s-OV}.

Since equation \eqref{eq:ds2s-OV} is rewritten as
\begin{gather*}
   \frac{x_k^{n+1}-x_k^n}{\delta t}
    = \delta x\Biggl\{
  {-}\frac{1}{\delta t}\biggl(\log\biggl[
      1+\biggl(\sum_{n^\prime=0}^{n_0}
      \dfrac{{\rm e}^{-(\Delta x_k^{n-n^\prime}-x_0-v_0\delta t)/\delta x}}{n_0+1}
      \biggr)^{-1}\biggr]\\
      \hphantom{\frac{x_k^{n+1}-x_k^n}{\delta t} =}{}
  -\log\biggl[
      1+\biggl(\sum_{n^\prime=0}^{n_0}
      \dfrac{{\rm e}^{-(\Delta x_k^{n-n^\prime}-x_0)/\delta x}}{n_0+1}
      \biggr)^{-1}\biggr]
  \biggr)\\
\hphantom{\frac{x_k^{n+1}-x_k^n}{\delta t} =}{}
 +\frac{\log\bigl(1+{\rm e}^{-(x_0+v_0\delta t)/\delta x}\bigr)
  + \log\bigl(1+{\rm e}^{-x_0/\delta x}\bigr)}{\delta t}\Biggr\} \\
\hphantom{\frac{x_k^{n+1}-x_k^n}{\delta t} }{}
 = v_0\biggl(1+\sum_{n^\prime=0}^{n_0}
      \dfrac{{\rm e}^{-(\Delta x_k^{n-n^\prime}-x_0)/\delta x}}{n_0+1}\biggr)^{-1}
      -v_0\bigl(1+{\rm e}^{x_0/\delta x}\bigr)^{-1}+O(\delta t),
\end{gather*}
the above dif\/ference equation~\eqref{eq:ds2s-OV} goes to an integral-dif\/ferential equation
in the continuum limit $\delta t\rightarrow 0$ as follows
\begin{gather}
  \frac{{\rm d} x_k}{{\rm d}t}
  = v_0\left(1+\dfrac{1}{t_0}\int_{0}^{t_0}
  {\rm e}^{-(\Delta x_k(t-t^\prime)-x_0)/\delta x}{\rm d}t^\prime
  \right)^{-1}
  -v_0\bigl(1+{\rm e}^{x_0/\delta x}\bigr)^{-1},
  \label{eq:s2s-OV}
\end{gather}
where $t_0:=n_0\delta t$,
$\frac{{\rm d}x_k}{{\rm d}t}
=\lim\limits_{\delta t\rightarrow 0}\frac{x_k^{n+1}-x_k^n}{\delta t}$ and
\begin{gather*}
    \lim_{\delta t\rightarrow 0}
    \sum_{n^\prime=0}^{n_0}
      \dfrac{{\rm e}^{-(\Delta x_k^{n-n^\prime}-x_0)/\delta x}}{n_0+1}
    = \dfrac{1}{t_0}\int_{0}^{t_0}
    {\rm e}^{-(\Delta x_k(t-t^\prime)-x_0)/\delta x}{\rm d}t^\prime.
\end{gather*}
In terms of an optimal velocity function and an ef\/fective distance
\begin{gather*}
  v_{\rm opt}(x)   := v_0\left(\frac{1}{1+{\rm e}^{-(x-x_0)/\delta x}}
  -\frac{1}{1+{\rm e}^{x_0/\delta x}}\right),\\
  \Delta_{\rm ef\/f}x_k(t)   := \delta x\log\left(
  \dfrac{1}{t_0}\int_{0}^{t_0}
  {\rm e}^{-\Delta x_k(t-t^\prime)/\delta x}{\rm d}t^\prime
  \right)^{-1},
\end{gather*}
the above integral-dif\/ferential equation is expressed as
\begin{gather*}
  \frac{{\rm d} x_k}{{\rm d} t} = v_{\rm opt}\bigl(\Delta_{\rm ef\/f}x_k(t)\bigr).
\end{gather*}
Since the ef\/fective
distance $\Delta_{\rm ef\/f}x_k(t)$ goes
to $\Delta x_k(t)$ in the limit below
\begin{gather*}
  \Delta x_k(t-t_0) = \lim_{h\rightarrow t_0}
  \delta x\log\left(
  \dfrac{1}{t_0-h}\int_{h}^{t_0}
  {\rm e}^{-\Delta x_k(t-t^\prime)/\delta x}{\rm d}t^\prime
  \right)^{-1},
\end{gather*}
this integral-dif\/ferential equation is an extension of
the Newell model \cite{Newell1961}
\begin{gather}
  \frac{{\rm d} x_k}{{\rm d} t}=v_{\rm opt}\bigl(\Delta x_k(t-t_0)\bigr),
  \label{eq:Newell}
\end{gather}
which is a car-following model dealing with retarded adaptation to the
optimal velocity determined by the headway in the past.

Replacement of $t$ with $t+t_0$ in equation~\eqref{eq:Newell} and the Taylor
expansion of $\dot{x}_k(t+t_0)=v_k(t+t_0)$ yield
\begin{gather*}
  v_{\rm opt}(\Delta x_k(t))   = v_k(t+t_0)
    = v_k(t)+\frac{{\rm d} v_k}{{\rm d} t}\cdot t_0
  +\frac{1}{2}\frac{{\rm d}^2 v_k}{{\rm d} t^2}\cdot t_0^2+\cdots,
\end{gather*}
which is equivalent to
\begin{gather}
  \frac{{\rm d} v_k}{{\rm d} t}
  +\frac{1}{2}\frac{{\rm d}^2 v_k}{{\rm d} t^2}\cdot t_0+\cdots
  =\frac{1}{t_0}\bigl(v_{\rm opt}(\Delta x_k(t))-v_k(t)\bigr).
  \label{eq:Taylor}
\end{gather}
The equation of motion of the OV model
\begin{gather*}
  \frac{{\rm d} v_k}{{\rm d} t}
  =\frac{1}{t_0}\bigl(v_{\rm opt}(\Delta x_k(t))-v_k(t)\bigr)
\end{gather*}
is given by neglecting the
higher order terms in the left hand side of the equation \eqref{eq:Taylor}.

The discussion shown above in this section shows
how the inverse ultradiscretization and the time continuous limit
connect the s2s model and the Newell model, which approximates
the OV model, through the s2s-OVCA.

\section{Flow-density relation}
\label{sec:3}

\begin{figure}[t]
\centering
\includegraphics[width=55mm]{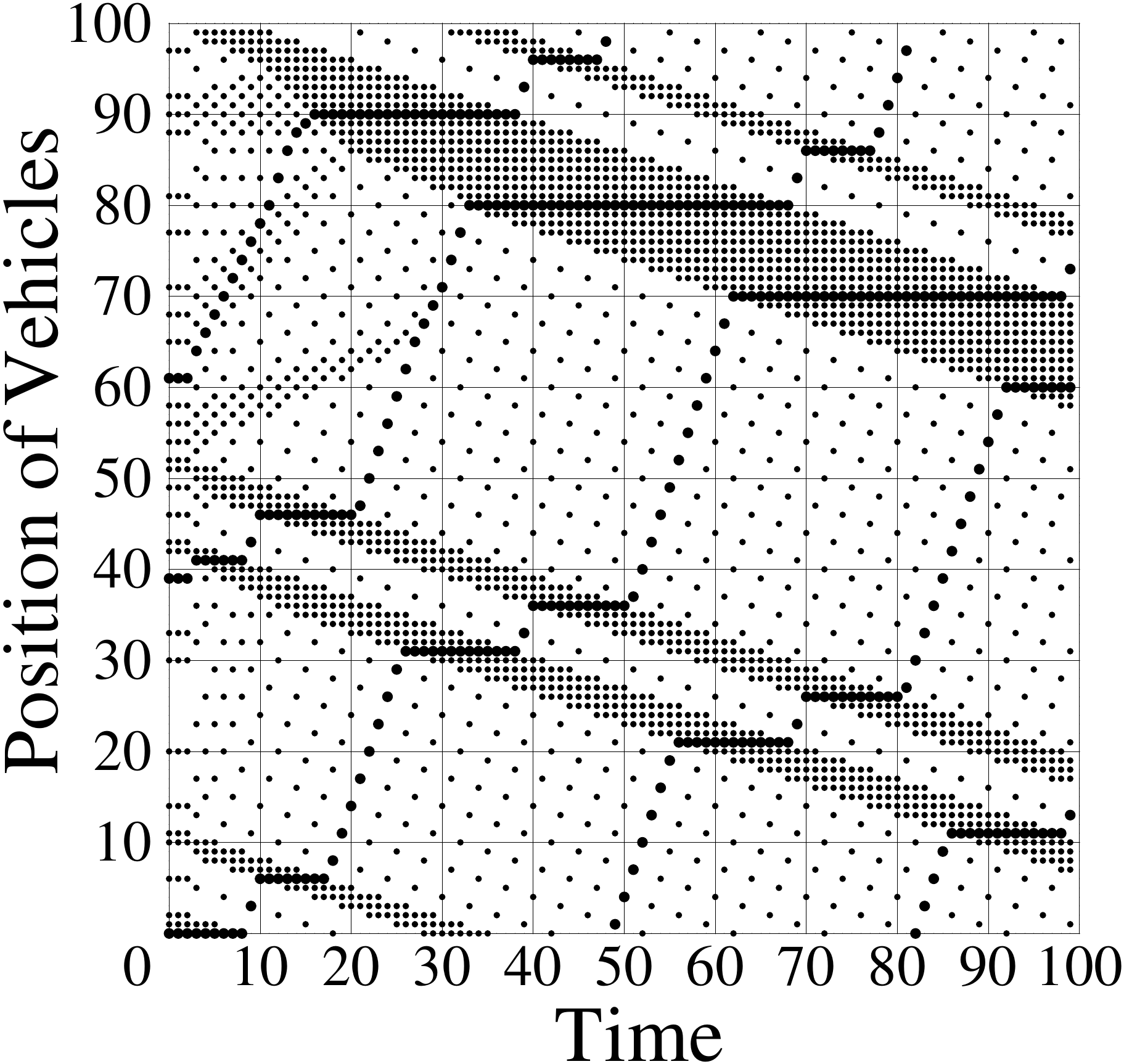}\quad
\includegraphics[width=53mm]{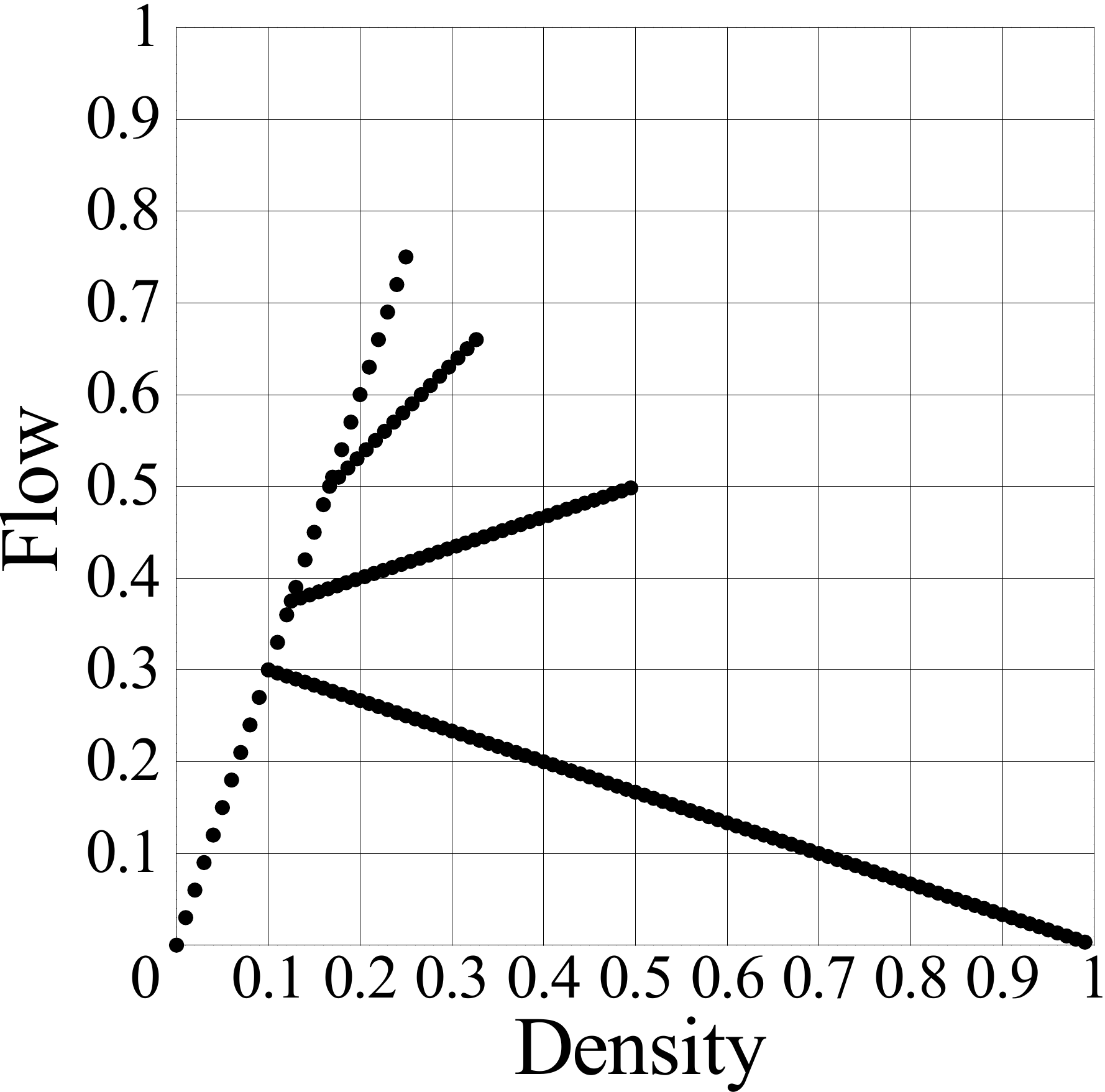}
\caption{The spatio-temporal pattern (left)
and the f\/low-density relation (right) of the s2s-OVCA~\cite{Oguma2009}.}
\label{fig:1}
\end{figure}

Fig.~\ref{fig:1} gives typical
examples of the spatio-temporal pattern showing jams and the
f\/low-density relation of the s2s-OVCA~\cite{Oguma2009}.
In the numerical calculation, the periodic boundary condition
is imposed and the length of the circuit $L$, which is the same
as the number of all the cells, is f\/ixed at $L=100$.
The maximum velocity $v_0$ and the monitoring
period $n_0$ are $v_0=3$ and $n_0=2$.
The number of the cars $K$ in the spatio-temporal pattern is set at $K=30$.

The spatio temporal pattern shows the trajectories of the cars.
As we can see, irregular motion of cars is observed in the early
stage of the time evolution, $0\leq n\leq 30$, where $n$ is the time.
But after that, the f\/low of the cars become stationary in the sense
that length of the jam is almost constant and that cars with
intermediate speeds appear only temporarily.

The f\/lows $Q$ in the f\/low-density relation are computed
by averaging over the time period $800=n_{\rm i}\leq n\leq n_{\rm f}= 1000$,
\begin{gather}
  Q:=\dfrac{1}{(n_f-n_i+1)L}\sum_{k=1}^K\sum_{n=n_i}^{n_f}v_k^n,
  \qquad v_k^n:=x_k^{n+1}-x_k^n,
  \label{eq:def_flow}
\end{gather}
in which the traf\/f\/ic is expected to be stationary
in the above mentioned sense.
The car density~$\rho$ is given by $\rho:=\frac{K}{L}$.
As we have mentioned before,
the f\/low-density relation of the s2s-OVCA is
piecewise linear and f\/lipped-$\lambda$ shaped with
several metastable slow branches.

The f\/low-density relation shown above is
derived by admitting the features of the f\/low
of the s2s-OVCA. Namely, the f\/low of the s2s-OVCA goes to one of the
stationary f\/lows in the long run. The stationary f\/lows consist of
the free f\/low in which all the cars run at the top speed $v_0$
and the slow f\/lows that
always contain slow cars running at the minimum speed $v_{\min}^\infty$,
$0\leq v_{\min}^\infty < v_0$, which remains constant. Formation of the
line of slow cars corresponds to that of traf\/f\/ic jam. In the slow f\/lows,
lengths of the jams are almost constant and f\/luctuate periodically.
Our previous paper~\cite{Ujino2012} gives a set of such stationary f\/lows.

\begin{figure}[t]
\centering
\includegraphics[width=55mm]{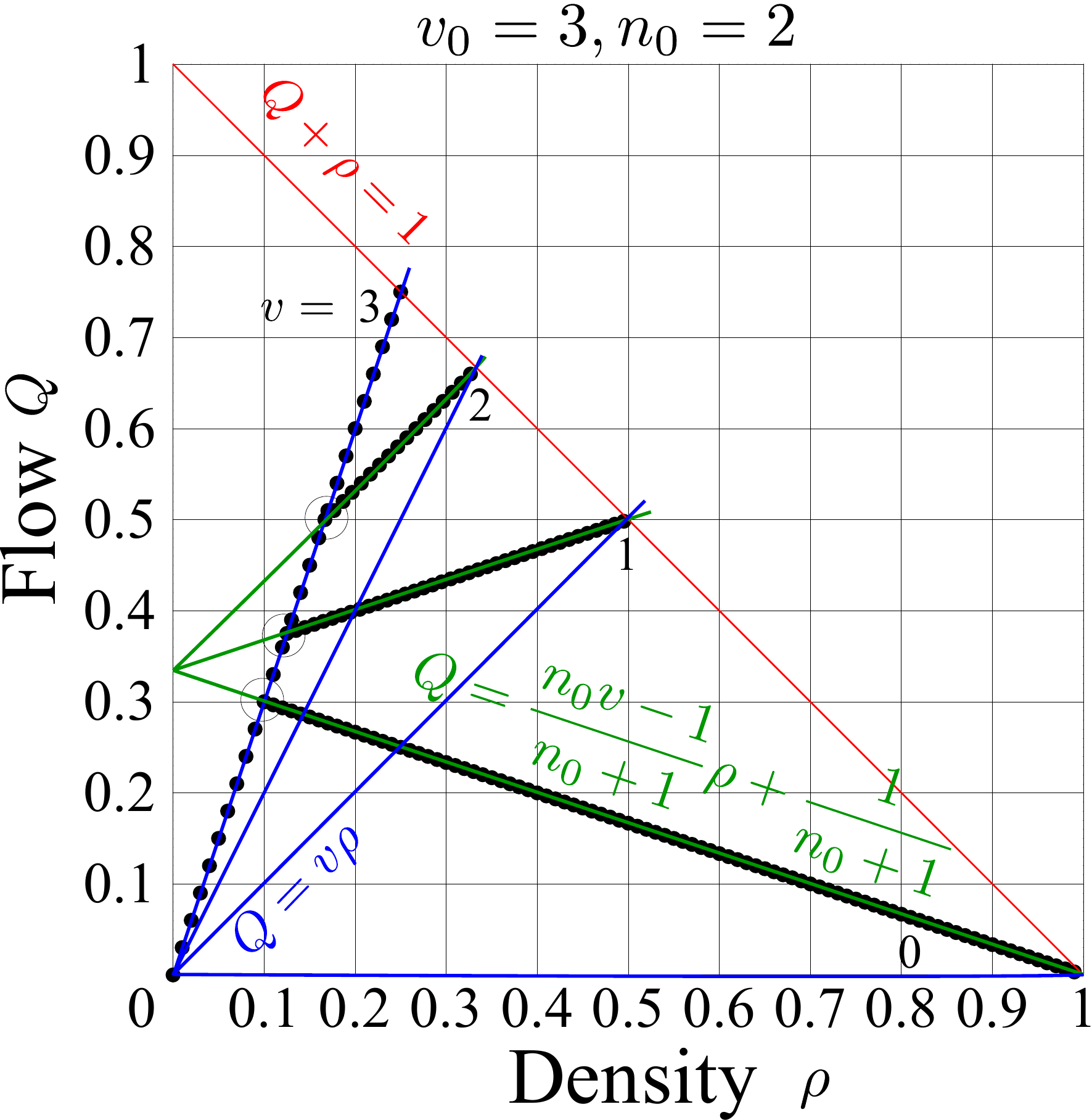}
\caption{The formula for the f\/low-density relation
of the s2s-OVCA~\cite{Ujino2012}.}
\label{fig:2}
\end{figure}

First we shall deal with the free f\/low and its f\/low-density relation.
Since all the cars run at the top speed, $v_k^n=v_0$, $\forall\, k$,
all the headways respectively
remain constant, $\Delta x_k^{n+1}=\Delta x_k^{n}\geq v_0+1$, $\forall \, k$.
Hence the f\/low~$Q$ is also constant in the future.
Using the def\/inition of the f\/low~\eqref{eq:def_flow}
with $n_{\rm i}=n_{\rm f}=n$, we have
\begin{gather*}
  Q=\frac{1}{L}\sum_{k=1}^K v_0=\rho v_0,
\end{gather*}
which gives the straight line in the f\/low-density
relation with a positive inclination that is equal to
the top speed $v_0$.
For example in Fig.~\ref{fig:2}, the f\/low-density relation of the
free f\/low which is labeled with $v=3$ is $Q=3\rho$, since $v_0=3$ in this case.

Next we shall consider the slow f\/lows and their f\/low-density relations.
Let us see a specif\/ic solution of the
s2s-OVCA starting from the following initial conf\/iguration
\begin{eqnarray}
  &0\colon & \verb*|1 2 3   4       5  6 7 8 9       0    |\,.
  \label{eq:example}
\end{eqnarray}
Note that the number 0 at the leftmost shows the time.
The digits and the blank symbols \verb*| | in the above conf\/iguration
mean the indices of the cars and the empty cells, respectively.
Thus the number of the cars~$K$ is~10 and
the length of the circuit~$L$ is~38 in this case.
We set the monitoring period $n_0$ at~2.
The speed of the cars 4, 9 and 0 is 3, which is the top speed~$v_0$
of this case. The speed of the car 5 is 2, whose headway is also~2.
All the other cars' speeds are~1, whose headways are also~1
except for the car~3. Thus the headways of the cars in tha past have
nothing to do with the motion of the cars in the future except for the car~3.
The headway of the car 3 at the time $-1$ is set to be~1.

Out of the above initial conf\/iguration~\eqref{eq:example},
the equation \eqref{eq:s2s-OVCA2} generates f\/low of vehicles
as follows
\begin{eqnarray*}
  &0\colon & \verb*|1 2 3   4       5  6 7 8 9       0    |\,, \\
  &1\colon & \verb*| 1 2 3     4      5 6 7 8   9       0 |\,, \\
  &2\colon & \verb*|0 1 2 3       4    5 6 7 8     9      |\,, \\
  &3\colon & \verb*| 0 1 2   3       4  5 6 7 8       9   |\,, \\
  &4\colon & \verb*|  0 1 2     3      4 5 6 7   8       9|\,, \\
  &5\colon & \verb*| 9 0 1 2       3    4 5 6 7     8     |\,, \\
  &6\colon & \verb*|  9 0 1   2       3  4 5 6 7       8  |\,.
\end{eqnarray*}
Note that the minimum speed of the cars $v_{\min}^\infty$ is 1
in the f\/low above.
We notice that the conf\/iguration at the time 3 is obtained by
moving all the cells of the initial conf\/iguration one cell
rightward as well as changing the car indices $k$ to $k-1$ modulo~10.
The conf\/iguration at the time 6 is also obtained by doing the same
shifts and changes of car indices to the conf\/iguration at the time~3.
In this sense, the above f\/low is a periodic motion of cars whose
period is 3 in this case. The length of the jam, or the number of the
cars running at the minimum speed, is thus almost constant.
Intermediate speeds also appear but only temporarily.
That is why we call them stationary
f\/lows of the s2s-OVCA. Roughly speaking, the slow f\/low
we shall deal with is the stationary f\/low of the type shown above.
The density of the cars~$\rho$ and the average f\/low $Q$
over the period, or the $n_0+1=3$ steps, are calculated as
\begin{gather*}
  \rho   = \frac{K}{L} =\frac{5}{19}, \\
  Q   = \frac{1}{(n_0+1)L}\sum_{k=0}^K
  \sum_{n^\prime=0}^{n_0}v_k^{n^\prime}
  =\frac{6+3+3+5+9+4+3+3+3+9}{3\times 38}=\frac{8}{19},
\end{gather*}
which will be verif\/ied with the formula we shall derive
shortly.

\begin{figure}[t]
\centering
\includegraphics[width=130mm]{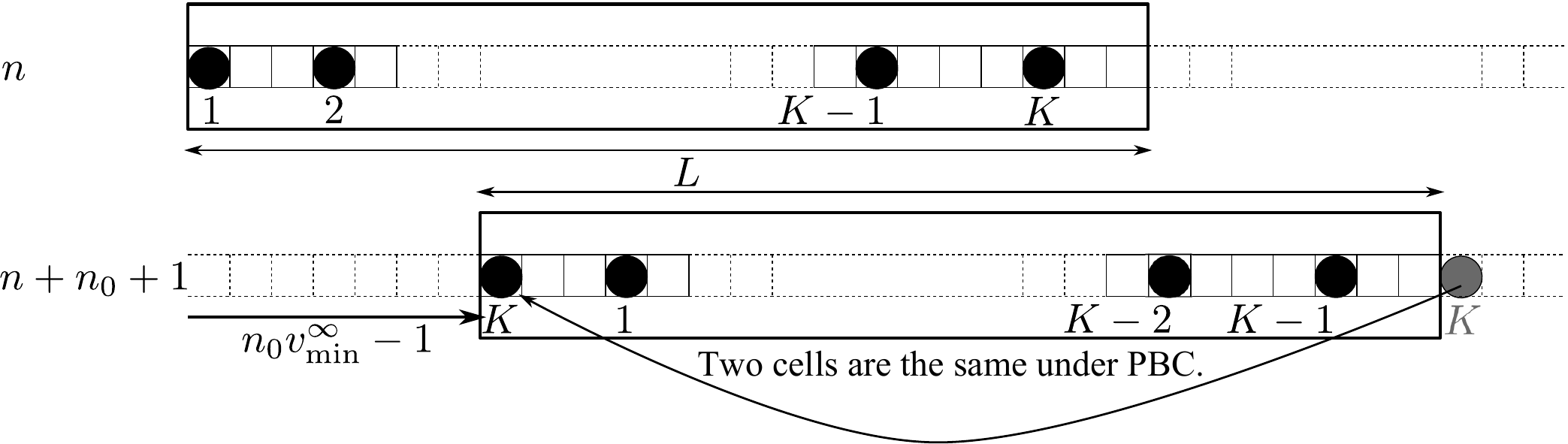}

\caption{The slow f\/low of the s2s-OVCA. Cars in the cells drawn by dashed
lines are omitted except for the car 1 at the time $n+n_0+1$.}
\label{fig:3}
\end{figure}

Let us consider such slow f\/lows as we have
seen above as the specif\/ic solutions in a more general manner.
Fig.~\ref{fig:3} shows conf\/igurations of a slow f\/low at times $n$
and $n+n_0+1$. Since we employ the periodic boundary condition,
two cells containing the car $K$ are identif\/ied. As a property of
the slow f\/low, we assume that the slow f\/low is periodic in the sense that
the conf\/iguration at the time $n+n_0+1$, which is shown in the box
in Fig.~\ref{fig:3},
is given by the rightward displacement of the
entire conf\/iguration at the time $n$ in the box
by $n_0v_{\min}^\infty-1$ cells.
The f\/low provided by this displacement of the entire
conf\/iguration in $n_0+1$ time steps is
$\frac{n_0 v_{\min}^\infty-1}{n_0+1}\rho$.
For example, the rightward displacement mentioned above for the
slow f\/low in Fig.~\ref{fig:3} is $2\times 1-1=1$, which agrees with
the observation before.
The set of stationary f\/lows given in~\cite{Ujino2012} has
the property of the slow f\/low we here assume.

Here we should note that
the leftward displacement of the car $K$ by $L$ cells, namely
whole the circuit length, which is f\/ictitiously introduced
to make the shifted initial conf\/iguration
from the real conf\/iguration  at the time $n_0+1$
in the sense that the numerical order
of the car arrays is maintained. In order to compensate the underestimation
of the f\/low brought about by this leftward displacement,
we have to add the f\/low corresponding to
the rightward displacement of the car $K$ by $L$ cells
in $n_0+1$ time steps, $\frac{1}{(n_0+1)L}\cdot L
=\frac{1}{n_0+1}$. Thus the f\/low of the slow f\/low with the minimum speed
$v_{\min}^\infty$ is given by
\begin{gather}
  Q=\frac{n_0 v_{\min}^\infty-1}{n_0+1}\rho+\frac{1}{n_0+1},\qquad
  0\leq v_{\min}^\infty<v_0.
  \label{eq:slow_lines}
\end{gather}
For example,
substitution of $\rho=\frac{5}{19}$, $n_0=2$ and
$v_{\min}^\infty=1$ into equation~\eqref{eq:slow_lines} yields
\begin{gather*}
  Q=\frac{2\times 1-1}{2+1}\times\frac{5}{19}+\frac{1}{2+1}
  =\frac{8}{19},
\end{gather*}
which agrees with the f\/low $Q=\frac{8}{19}$
for the slow f\/low given above as an specif\/ic solution.
The formula~\eqref{eq:slow_lines}
agrees with the f\/low-density relation given by numerical experiments,
as we can see in Fig.~\ref{fig:2}.
Three branches labeled with $v=2, 1$ and $0$
are the f\/low-density relations with
the minimum speeds $v_{\min}^\infty=v$ in Fig.~\ref{fig:2}.
Roughly speaking,
the traf\/f\/ic that forms the branch corresponding to $v_{\min}^\infty$
consists of groups of cars running at the top speed $v_0$ and other
groups of cars running at the minimum speed $v_{\min}^\infty$,
as the specif\/ic solution evolving from the initial conf\/iguration~\eqref{eq:example} has shown. A set of specif\/ic solutions that
correspond to the branches in the fundamental diagram is given
in~\cite{Ujino2012}.

The maximum
density $\rho_{\max}(v_{\min}^\infty)$
that allows the minimum speed to be $v_{\min}^\infty$ is
\begin{gather}
  \rho_{\max}(v_{\min}^\infty)
  =\frac{1}{v_{\min}^\infty+1}.
  \label{eq:9}
\end{gather}
The f\/low $Q(\rho_{\max}(v_{\min}^\infty))$ corresponding to
the maximum density $\rho_{\max}(v_{\min}^\infty)$ is then given by
\begin{gather}
  Q(\rho_{\max}(v_{\min}^\infty))
  =\rho_{\max}(v_{\min}^\infty)v_{\min}^\infty.
  \label{eq:10}
\end{gather}
Since the two equations~\eqref{eq:9} and~\eqref{eq:10} holds at the same time,
they leads to
$Q(\rho_{\max}(v_{\min}^\infty))+\rho_{\max}(v_{\min}^\infty)=1$.
Thus all the end points of the branches must be on the line
\begin{gather}
  Q+\rho=1.
  \label{eq:max_rho}
\end{gather}
The branching point, or the minimum density, of the f\/low-density relation
of the slow f\/low corresponding to the minimum speed $v_{\min}^\infty$
is determined by the intersection of the f\/low density relations of
the free f\/low and the slow f\/low
\begin{gather}
  \rho_{\min}(v_{\min}^\infty)
  =\frac{1}{n_0(v_0-v_{\min}^\infty)+v_0+1}.
  \label{eq:min_rho}
\end{gather}
In Fig.~\ref{fig:2}, the branching points corresponding to
$v_{\min}^\infty=2, 1$ and 0 are encircled with small circles,
which agree with the above formula~\eqref{eq:min_rho}.
The density of the cars $\rho$ needs to be suf\/f\/iciently large
so as to form the slow f\/low with the minimum speed
$v_{\min}^\infty$. The branching point gives the lower bound of
such density.

Since the s2s-OVCA \eqref{eq:s2s-OVCA2} is a deterministic CA,
the initial conf\/iguration determines the f\/inal state. Its numerical simulation
is very robust against, or more precisely speaking, free from numerical errors.
Thus all the stationary f\/lows including the free f\/low and the slow f\/lows beyond
the minimum density $\rho_{\min}(v_{\min}^\infty)$ are stable in the
numerical simulation. That is why we were able to obtain
the f\/low-density relation with several ``metastable'' states
consisted by the free f\/low and the slow f\/lows beyond
the minimum density, as was shown in Figs.~\ref{fig:1} and~\ref{fig:2}.
Though these metastable states are robust against numerical errors in simulation, but they are generally unstable against perturbation, which gives
a reason of their name. For example, when one gives a perturbation
to the headways that is equal to the minimum
speed $v_{\min}^\infty$ in the groups of cars running at the minimum
speed $v_{\min}^\infty$, a car with a velocity that is less
than $v_{\min}^\infty$ appears.
And such a perturbed car generally becomes a seed of a group of
cars running at a
velocity slower than $v_{\min}^\infty$, which eventually slows
down whole the traf\/f\/ic. That is why we call these states metastable,
except for the slow line with $v_{\min}^\infty=0$,
which we cannot make slower.

When the monitoring period $n_0$ is zero, all of the slow f\/lows
\eqref{eq:slow_lines} goes to the line of the end points of the metastable
branches \eqref{eq:max_rho}.
Thus the monitoring period plays an essential role
in the formation of the metastable states in the f\/low-density relation.
Thus it could be determined by comparing the f\/low-density relation of the
s2s-OVCA and observed ones.

\section{Summary}
We have shown an inverse ultradiscretization from the
s2s-OVCA~\eqref{eq:s2s-OVCA2}
to an integral-dif\/ferential equation~\eqref{eq:s2s-OV}, which is an extension
of the Newell model~\eqref{eq:Newell}.
Since the Newell model~\cite{Newell1961} and the s2s-OVCA~\cite{Oguma2009}
are extended models
of the OV~\cite{Bando1995} and the s2s models~\cite{Takayasu1993} respectively,
the s2s-OVCA is interpreted as a CA-type hybrid of the OV and the s2s models.

Using the features of the stationary f\/lows observed
in the numerical experiments, we have derived the f\/low-density relations
of the stationary f\/low of the s2s-OVCA. The f\/low-density relations of
the s2s-OVCA were numerically obtained~\cite{Oguma2009}
and then derived by use of a set of stationary f\/lows~\cite{Ujino2012}.

Since the s2s-OVCA is a deterministic CA,
the model is suitable for a simulation with a~much bigger system size
than the length of the circuit, $L=100$, in our numerical simulation.
We expected that it would be suf\/f\/icient to capture the
characteristics of stationary f\/lows of the s2s-OVCA on the circuit,
which is determined by the density of cars and initial conf\/iguration.
In order to observe f\/inite-size ef\/fects of open boundaries, for example,
we expect that the s2s-OVCA will be a good tool.

The s2s-OVCA has several types of monotonicity in its time evolution,
which extend the results shown for the $n_0=1$
case~\cite{Tian2009}. We expect that the monotonicity determines the
relaxation to the stationary f\/low from the initial conf\/iguration as well as
the property of the stationary f\/low we assume here.
We hope that results
on the relaxation to stationary f\/lows and the monotonicity
in the time evolution of the s2s-OVCA will be reported soon.

\subsection*{Acknowledgments}

One of the authors (HU) is grateful to K.~Oguma for the previous collaboration.

\pdfbookmark[1]{References}{ref}
\LastPageEnding


\begin{thebibliography}{99}
\footnotesize \itemsep=0pt

\bibitem{Bando1995}
Bando M., Hasebe K., Nakayama A., Shibata A., Sugiyama Y., Dynamical model of
  traf\/f\/ic congestion and numerical simulation, \href{http://dx.doi.org/10.1103/PhysRevE.51.1035}{\textit{Phys. Rev.~E}}
  \textbf{51} (1995), 1035--1042.

\bibitem{Barlovic1998}
Barlovic R., Santen L., Schadschneider A., Schreckenberg M., Metastable states
  in cellular automata for traf\/f\/ic f\/low, \href{http://dx.doi.org/10.1007/s100510050504}{\textit{Eur. Phys.~J.~B}} \textbf{5}
  (1998), 793--800, \href{http://arxiv.org/abs/cond-mat/9804170}{cond-mat/9804170}.

\bibitem{Chowdhury2000}
Chowdhury D., Santen L., Schadschneider A., Statistical physics of vehicular
  traf\/f\/ic and some related systems, \href{http://dx.doi.org/10.1016/S0370-1573(99)00117-9}{\textit{Phys. Rep.}} \textbf{329} (2000),
  199--329, \href{http://arxiv.org/abs/cond-mat/0007053}{cond-mat/0007053}.

\bibitem{Fukui1996}
Fukui M., Ishibashi Y., Traf\/f\/ic f\/low in 1D cellular automaton model including
  cars moving with high speed, \href{http://dx.doi.org/10.1143/JPSJ.65.1868}{\textit{J.~Phys. Soc. Japan}} \textbf{65} (1996),
  1868--1870.

\bibitem{Helbing2001}
Helbing D., Traf\/f\/ic and related self-driven many-particle systems, \href{http://dx.doi.org/10.1103/RevModPhys.73.1067}{\textit{Rev.
  Modern Phys.}} \textbf{73} (2001), 1067--1141, \href{http://arxiv.org/abs/cond-mat/0012229}{cond-mat/0012229}.

\bibitem{Helbing1999}
Helbing D., Schreckenberg M., Cellular automata simulating experimental
  properties of traf\/f\/ic f\/low, \href{http://dx.doi.org/10.1103/PhysRevE.59.R2505}{\textit{Phys. Rev.~E}} \textbf{59} (1999),
  R2505--R2508, \href{http://arxiv.org/abs/cond-mat/9812300}{cond-mat/9812300}.

\bibitem{Kanai2009}
Kanai M., Isojima S., Nishinari K., Tokihiro T., Ultradiscrete optimal velocity
  model: a~cellular-automaton model for traf\/f\/ic f\/los and linear instability of
  high-f\/lux traf\/f\/ic, \href{http://dx.doi.org/10.1103/PhysRevE.79.056108}{\textit{Phys. Rev.~E}} \textbf{79} (2009), 056108, 8~pages,
  \href{http://arxiv.org/abs/0902.2633}{arXiv:0902.2633}.

\bibitem{Newell1961}
Newell G.F., Nonlinear ef\/fects in the dynamics of car f\/lowing,
  \href{http://dx.doi.org/10.1287/opre.9.2.209}{\textit{Operations Res.}} \textbf{9} (1961), 209--229.

\bibitem{Oguma2009}
Oguma K., Ujino H., A hybrid of the optimal velocity and the slow-to-start
  models and its ultradiscretization, \href{http://dx.doi.org/10.14495/jsiaml.1.68}{\textit{JSIAM Lett.}} \textbf{1} (2009),
  68--71, \href{http://arxiv.org/abs/0908.3377}{arXiv:0908.3377}.

\bibitem{Takahashi2009}
Takahashi D., Matsukidaira J., On a discrete optimal velocity model and its
  continuous and ultradiscrete relatives, \href{http://doi.org/10.14495/jsiaml.1.1}{\textit{JSIAM Lett.}} \textbf{1}
  (2009), 1--4, \href{http://arxiv.org/abs/0809.1265}{arXiv:0809.1265}.

\bibitem{Takahashi1990}
Takahashi D., Satsuma J., A soliton cellular automaton, \href{http://dx.doi.org/10.1143/JPSJ.59.3514}{\textit{J.~Phys. Soc.
  Japan}} \textbf{59} (1990), 3514--3519.

\bibitem{Takayasu1993}
Takayasu M., Takayasu H., $1/f$ noise in a traf\/f\/ic model, \href{http://dx.doi.org/10.1142/S0218348X93000885}{\textit{Fractals}}
  \textbf{1} (1993), 860--866.

\bibitem{Tian2009}
Tian R., The mathematical solution of a cellular automaton model which
  simulates traf\/f\/ic f\/low with a~slow-to-start ef\/fect, \href{http://dx.doi.org/10.1016/j.dam.2009.03.013}{\textit{Discrete Appl.
  Math.}} \textbf{157} (2009), 2904--2917.

\bibitem{Tokihiro1996}
Tokihiro T., Takahashi D., Matsukidaira J., Satsuma J., From soliton equations
  to integrable cellular auto\-mata through a limiting procedure, \href{http://dx.doi.org/10.1103/PhysRevLett.76.3247}{\textit{Phys.
  Rev. Lett.}} \textbf{76} (1996), 3247--3250.

\bibitem{Ujino2012}
Ujino H., Yajima T., Exact solutions and f\/low-density relations for a cellular
  automaton variant of the optimal velocity model with the slow-to-start
  ef\/fect, \href{http://dx.doi.org/10.1143/JPSJ.81.124005}{\textit{J.~Phys. Soc. Japan}} \textbf{81} (2012), 124005, 8~pages,
  \href{http://arxiv.org/abs/1210.7562}{arXiv:1210.7562}.

\bibitem{Wolfram1986}
Wolfram S. (Editor), Theory and applications of cellular automata,
  \textit{Advanced Series on Complex Systems}, Vol.~1, World Scientif\/ic
  Publishing Co., Singapore, 1986.

\end{thebibliography}
\end{document}